%% file: example_paper.tex
\documentclass{article}

\usepackage{microtype}
\usepackage{graphicx}
\usepackage{subfigure}
\usepackage{booktabs} %

\usepackage{hyperref}

\usepackage[accepted]{icml2023}

\usepackage{amsmath}
\usepackage{amssymb}
\usepackage{mathtools}
\usepackage{amsthm}

\usepackage[capitalize,noabbrev]{cleveref}

\usepackage{xspace}
\usepackage{pgf-pie}
\usepackage{physics}
\usepackage{multirow}

\theoremstyle{plain}

\theoremstyle{definition}

\theoremstyle{remark}

\usepackage[textsize=tiny]{todonotes}

\icmltitlerunning{Computational Approaches for App-to-App Retrieval and Design Consistency Check}

\begin{document}

\twocolumn[
\icmltitle{Computational Approaches for App-to-App Retrieval \\ 
and Design Consistency Check}

\icmlsetsymbol{equal}{*}

\begin{icmlauthorlist}
\icmlauthor{Seokhyeon Park}{equal,snu}
\icmlauthor{Wonjae Kim}{equal,naver}
\icmlauthor{Young-Ho Kim}{naver}
\icmlauthor{Jinwook Seo}{snu}
\end{icmlauthorlist}

\icmlaffiliation{naver}{Naver AI Lab, Seongnam, Gyeonggi, Republic of Korea}
\icmlaffiliation{snu}{Department of Compuster Science and Engineering, Seoul National University, Seoul, Republic of Korea}

\icmlcorrespondingauthor{Jinwook Seo}{jseo@snu.ac.kr}

\icmlkeywords{Machine Learning, ICML}

\vskip 0.3in
]

\printAffiliationsAndNotice{\icmlEqualContribution} %

\input{Sections/lib}

\def\scrt{screenshot-to-screenshot\xspace}
\def\sett{set-to-set\xspace}
\def\appt{app-to-app\xspace}
\def\dot{$\mathcal{D}_{ot}$\xspace}
\def\lossu{$L_u$\xspace}
\def\deltalu{$\Delta L_u$\xspace}

\begin{abstract}

\input{Sections/0-Abstract}

\end{abstract}

\section{Introduction}
\input{Sections/1-Introduction}

\section{Background}
\input{Sections/2-Background}

\section{Methods}
\input{Sections/3-Methods}

\section{Experiments}
\input{Sections/4-Experiments}

\section{Limitations and Future Work}
\input{Sections/5-Limitation}

\section{Conclusion}
\input{Sections/6-Conclusion}

\bibliography{example_paper}
\bibliographystyle{icml2023}

\end{document}

%% file: Sections/lib.tex
\newcommand{\eg}{\textit{e.g.}}
\newcommand{\ie}{\textit{i.e.}}
\newcommand{\cf}{\textit{c.f.}}
\newcommand{\etal}{\textit{et al.}}

\definecolor{revisedcolor}{RGB}{0,0,255}

\newenvironment{notready}{\color{gray}}{}
\newcommand{\needtorevise}[1]{\textcolor{blue}{#1}}
\newcommand{\needtocheck}[1]{\textcolor{red}{#1}}
\newcommand{\revised}[1]{\textcolor{blue}{#1}}

\newcommand{\cameraready}[1]{#1}

\newcommand{\ipstart}[1]{\vspace{1mm} \noindent{\textbf{\textit{#1.}}}}

\newcommand{\circledigit}[1]{\textbf{\normalsize{\textsf{\textcircled{\footnotesize{#1}}}}}}

\definecolor{tableheader}{HTML}{EFEFEF}
\definecolor{tablegrayline}{HTML}{d0d0d0}

\definecolor{darkgray}{HTML}{555555}

%% file: Sections/0-Abstract.tex
Extracting semantic representations from mobile user interfaces (UI) and using the representations for designers' decision-making processes have shown the potential to be effective computational design support tools.
Current approaches rely on machine learning models trained on small-sized mobile UI datasets to extract semantic vectors and use screenshot-to-screenshot comparison to retrieve similar-looking UIs given query screenshots.
However, the usability of these methods is limited because they are often not open-sourced and have complex training pipelines for practitioners to follow, and are unable to perform screenshot set-to-set (\ie, app-to-app) retrieval.
To this end, we (1) employ visual models trained with large web-scale images and test whether they could extract a UI representation in a zero-shot way and outperform existing specialized models, and (2) use mathematically founded methods to enable app-to-app retrieval and design consistency analysis.
Our experiments show that our methods not only improve upon previous retrieval models but also enable multiple new applications.

%% file: Sections/1-Introduction.tex
Designing attractive yet informative user interfaces for mobile applications is as important, if not more, as engineering the core functionality of the app.
In the early phase of the design process, designers often rely on curation services such as Dribbble or Pinterest to get inspiration \citep{wu2021exploring, lu2022bridging} and search for the design choices of their competitor apps.
As the app reaches its release, designers focus on validating the overall consistency of their app to maximize the quality of the app, rather than seeking more inspiration. \citep{lu2022bridging}
For senior-level designers, it is sometimes more important to maintain the design consistency among different screens of their app than to consider design alternatives of the curation services \citep{wu2021exploring}.

To help the inspiration process, the HCI and Machine Learning communities have been trying to model the process as a visual search task, which is essentially a UI \scrt retrieval task \citep{kumar2013webzeitgeist,ritchie2011d,huang2019swire,chen2020wireframe,li2021screen2vec,bunian2021vins,liu2018learning}.
Using machine learning-based models, recent works such as Screen2Vec \citep{li2021screen2vec} or VINS \citep{bunian2021vins} first embedded the UI screenshots and their accompanying metadata including UI view hierarchy \citep{deka2017rico} into vectors representing the semantics of the screenshots.
As the distance between these encoded vectors indicates the similarity between the original screenshots, designers can easily query similar UI screenshots to an input.

Designers in the field often design UIs as \textit{UI flow} to support the users' interaction flow, which involves multiple UI screenshots\footnote{In this paper, we use the term \textit{screenshot} to denote a rasterized image of an app screen without any metadata about the screen elements}.
However, existing systems are bound to \scrt retrieval based on a single UI screenshot, which creates challenges when the designers want to explore design alternatives for a sequence of UI screenshots, (\eg, ``\textit{which app contains UI flows that are semantically similar to ours?}''), such models hardly support these tasks due to the lack of well-defined measures for set-to-set distance.
Building upon the concept of \scrt retrieval, we propose a novel \appt retrieval using a mathematically founded optimal transport method.
The method not only provides a scalar metric that represents the distance between the apps but also shows the optimal matching score between UI pages in the queried and retrieved apps, enhancing the interpretability of the system.
Since most of the apps in the widely used Rico dataset \citep{deka2017rico} contain less than 10 screenshots, it is challenging to show the usefulness of \appt retrieval.
Consequently, to show the effectiveness of our proposed method, we gathered a new dataset from Mobbin\footnote{\url{https://mobbin.com/}}, a curated mobile UI screenshot hub, where each app includes 126 screenshots on average.

Another crucial task enabled by the analysis of set-level UI representation is the automated validation of the UI design consistency, a domain that has yet to be extensively explored.
Early studies \citep{mahajan1997visual, ivory2001state} along with recent ones \citep{yang2021don,burny2022semi} attempted to predict whether queried UIs violated heuristic design guidelines.
While valuable, these guidelines often prove insufficient as field designers prioritize the company's own guidelines.
Consequently, academic guidelines are often dismissed as they might conflict with design intentions \citep{colusso2017translational}, suggesting a promising direction for future research.

We posit that it is vital to employ data-driven design rules directly from designers' queries (\ie, pre-existing UIs in the app) without resorting to heuristics.
By doing so, the \textit{extracted} semantics can be more closely aligned with the designers' intention.
To achieve such a goal, we exploit the metric of \textit{uniformity} \citep{wang2020understanding} to measure the consistency among UIs in a specific set (\ie, UIs in the app).
With the uniformity metric, practitioners would easily measure the effect of newly added sets of UIs and compare different alternatives.

For both retrieval and consistency check tasks, it is required to acquire the \textit{semantic representation} of the graphical UI.
Recent works \citep{li2021screen2vec,bunian2021vins} employ machine learning (ML) models to produce semantic vectors from UI screens, typically trained on small-sized datasets like Rico \citep{deka2017rico}.
However, advancements in foundation models \citep{bommasani2021opportunities} demonstrate that models trained on extensive web-scale datasets can surpass those trained on smaller, meticulously curated datasets, as evidenced by the success of GPT-3 \citep{brown2020language} and CLIP \citep{radford2021learning}.

Since it is still unclear whether models trained on uncurated web-crawled datasets can outperform specialized models trained for UI retrieval using well-curated UI screens, we conducted qualitative and quantitative studies to evaluate our model.
As a result, we found that the model actually grasps the semantics of UIs in a zero-shot way and its semantic vectors result in more preferred UI retrieval compared to specialized models when tested with human crowd workers.

We summarize our contributions as follows:
\begin{itemize}
    \item We extend \scrt retrieval to \appt retrieval with an optimal transport method, enabling new applications of the data-driven UI design inspiration process.
    \item We rethink the task of design consistency check and enable data-driven consistency check with the metric of uniformity.
    \item We investigate the zero-shot applicability of visual foundation models for UI semantics through comprehensive experiments.
\end{itemize}

%% file: Sections/2-Background.tex
In this section, we cover related work in the areas of (1) computational UI understanding and its adoption; (2) optimal transport for an app-to-app retrieval; and (3) uniformity for design consistency check.

\subsection{Computational UI understanding and its adoption}

Multiple studies \citep{bonnardel1999creativity,wu2021exploring,lu2022bridging,herring2009getting} report that designers prefer to get inspiration from pre-existing design examples.
\citet{lee2010designing} showed that designers use various examples during the design process, and even showed that interfaces created through the process involving multiple references are preferred over those without references.
To meet such needs, the HCI community has studied computational approaches for understanding user interfaces \citep{jiang2022computational}.
From traditional computer vision algorithms \citep{kumar2013webzeitgeist,ritchie2011d} to deep learning models \citep{huang2019swire,chen2020wireframe,li2021screen2vec,bunian2021vins,liu2018learning}, the community tried to distill the semantics of given UI screenshots into usable representations.

Although beneficial \textit{prima facie}, practitioners are frustrated with these tools for multiple reasons \citep{lu2022bridging,colusso2017translational}.
As \citet{lu2022bridging} pointed out, one of the most common pitfalls of these models is their lack of ability to understand an app's problem domains and functionalities.
The problem arises as these models only allow queries with a single UI screenshot, and it's nearly impossible to infer the app's intention with a single screenshot. In this work, we seek ways to support querying and retrieving a similar \textit{app} instead of a single UI screenshot.

\subsection{Optimal transport for an app-to-app retrieval}
\begin{figure}[t]
\begin{center}
    \centerline{\includegraphics[width=1.1\linewidth]{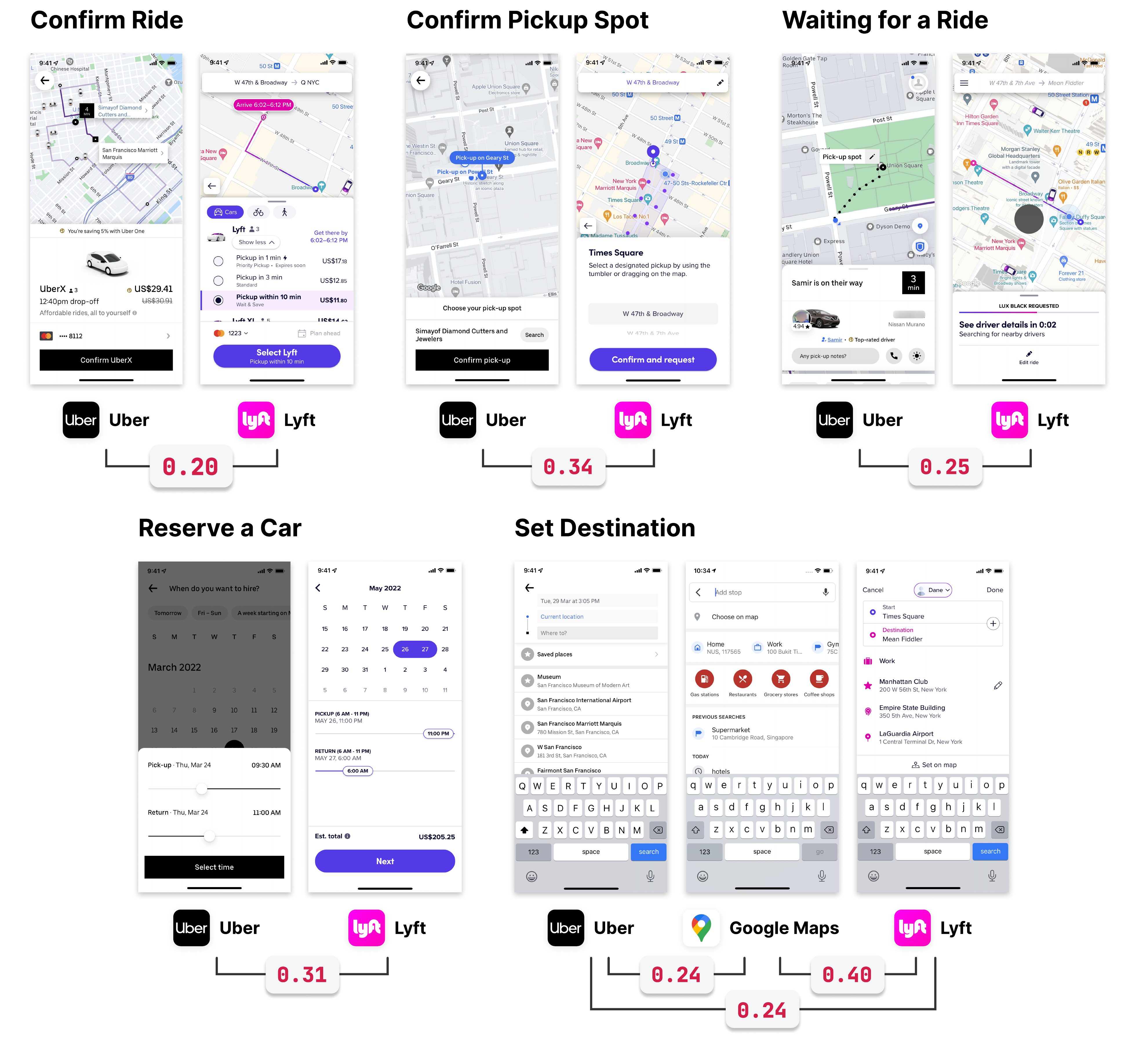}}
    \vspace{-0.5em}
    \caption{UI screenshots of various functionality from the Uber, Lyft, and Google Maps apps. The red number is a cosine distance between the representations of two screenshots. For \textit{set destination} UI screenshot, we can see the distances are the same between the Uber and Google Maps pair and the Uber and Lyft pair.}
    \label{fig:uber_lyft}
\end{center}
\vspace{-3em}
\end{figure}

Defining an app as a set of UI screenshots, we can derive an app's problem domains and functionalities from the relations between screenshots from different apps. For example, in \cref{fig:uber_lyft}, if designers use the \textit{set destination} UI screenshot of the Uber app, its cosine distance to Google Maps and Lyft app's set destination screenshot is equal.
However, considering screens for other functionalities of the Uber app (\eg, \textit{confirm ride}, \textit{reserve car}, \dots) comprehensively, we can more reliably infer that Lyft is semantically closer to Uber than Google Maps.
This process is revisited later in \cref{fig:app_retrieval}.

Comparing multiple screenshots congruently can be viewed as the transportation of virtual masses from a query set of UIs to a target set of UIs.
As deep-learning-based models make UIs into vectors on the n-dimensional Euclidean space $\mathbb{R}^{n}$, we can imagine virtual masses distributed over the set of vectors of the app.
Then, there exist multiple transportation plans that transport these masses to the target app's UI vectors, and each plan can be written as a doubly stochastic matrix as it should preserve the total amount of the masses.
Considering both the transportation plan and the distance between each pair, we can compute the optimal transport (OT) plan \citep{villani2009optimal,peyre2019computational,santambrogio2015optimal}, which minimizes the transportation cost.
Such cost is a scalar value that describes the distance between two apps and can be used to enable app-to-app retrieval. In this work, we leverage the OT plan to efficiently compute the relationship between apps to boost up both the latency and quality of the retrieval task.

\subsection{Uniformity for design consistency check}
\label{sec:uniformity_background}
Beyond inspiration, checking the overall consistency of screens for the same app (\eg, checking the consistency of a new design draft against the existing screens) is also one of the primary tasks for app designers \citep{wu2021exploring}.
Traditionally, the HCI community has focused on producing general guidelines predominantly targeted for inclusiveness (\eg, accessibility) and detecting violations of the interface guidelines in an automated manner \citep{mahajan1997visual, ivory2001state,yang2021don,burny2022semi}.
Despite the usefulness of guidelines, it is not easy for industrial designers to comply with the guidelines because they might already have the company's own design guidelines, not to mention they have to reinterpret and adapt them to their situation.
In this work, we aim to handle this gap by measuring the consistency of an app and treating the violation as a decrement in consistency.
Although our approaches do not provide explicitly documented guidelines, it is generally applicable to all situations as long as the reference set of UI screens (\ie, app) exists.

Given a set of vectors, we can consider the Gaussian potential kernel that maps a pair of vectors into a positive scalar.
If we normalize vectors onto an (n-1)-dimensional unit hypersphere $\mathcal{S}^{n-1}$, the distributions of vectors minimizing the average pairwise Gaussian potential (i.e. uniformity) weakly converge to the uniform distribution \citep{wang2020understanding}.
That says, if we use contrastive representation models such as CLIP \citep{radford2021learning}, we can measure how uniformly the vectors in the set are distributed, as contrastive loss contains the term minimizing uniformity.
This property can directly be used for measuring the consistency of the app as an app with less uniformity means an app consists of similar UIs, thus high consistency.

%% file: Sections/3-Methods.tex
In this section, we describe our methodologies for collecting a mobile app screenshot dataset, calculating vector representations from screenshots, and the optimal transport to implement an app-to-app retrieval task.

\subsection{Dataset}

\begin{figure}[t]
\begin{center}
    \centerline{\includegraphics[width=\linewidth]{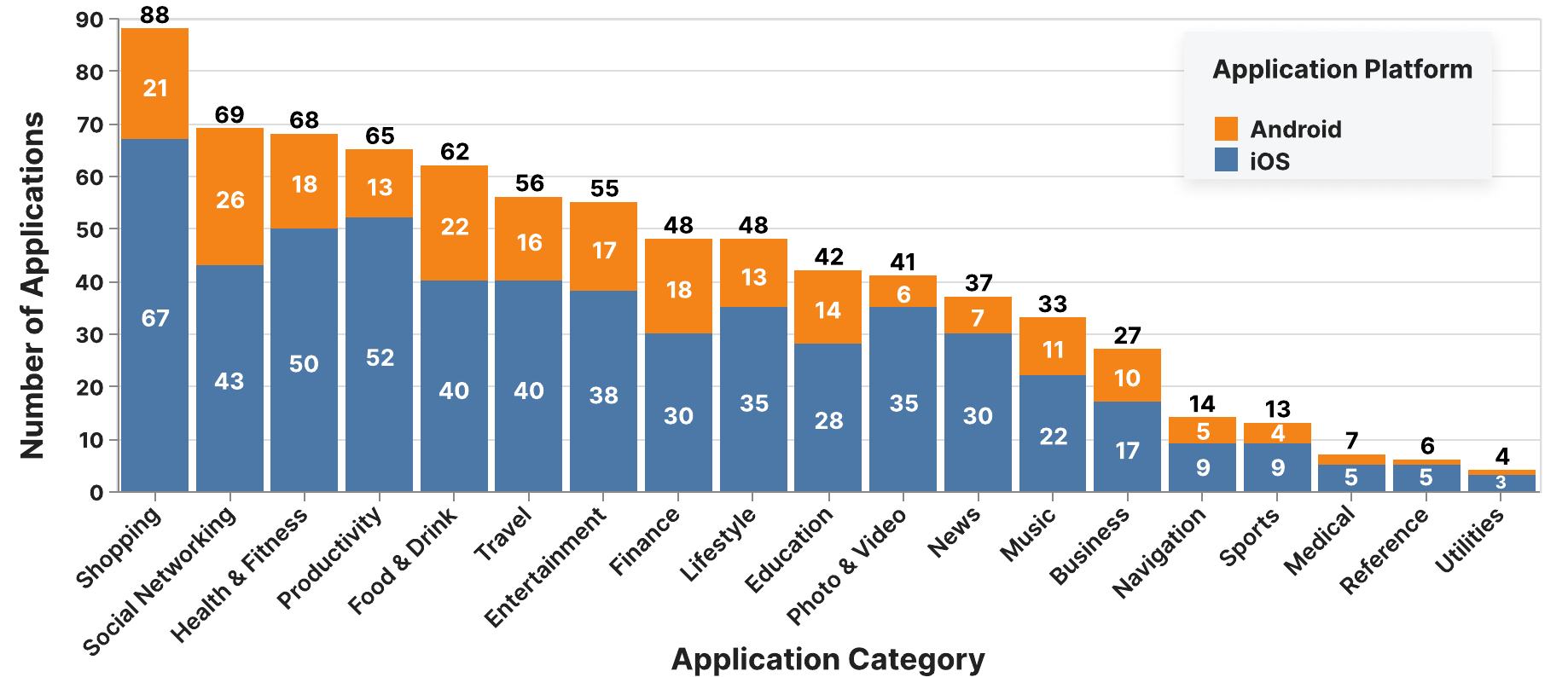}}
    \vspace{-1em}

    \caption{Number of applications in Mobbin dataset by app category and platform}
    \label{fig:dataset_chart}
\end{center}
\vspace{-3.5em}
\end{figure}

A variety of mobile UI screenshot datasets have been proposed as the basis for data-driven interface studies.
However, these datasets are mostly devised for screenshot-level analysis.
For example, the widely used Rico dataset \citep{deka2017rico} is limited to being used for app-level analysis as more than 75\% of apps in Rico have less than 10 screens.
Other datasets also have limitations to be used for our analysis such as the public availability and size of the datasets.

To properly evaluate our methods of app-level UI analysis, we collected a new dataset from Mobbin, which is a UI curation service providing up-to-date app-wise screenshots.
The dataset was obtained as of June 2022 and has a total of 320 unique mobile apps, and each app has its own sets of screenshots based on the platform (iOS or Android) and the app version (date), which makes a total of 783 sets of screenshots, 558 for iOS and 225 for Android.
It has a total of 99,228 screenshots (62,315 for iOS and 36,913 for Android), and each app has an average of 126 screenshots, which is significantly larger than both Rico's total screenshot counts and the average number of screenshots per app.
The Mobbin dataset consists of apps with 19 categories, and the number of apps for each category and the platform is shown in \cref{fig:dataset_chart}.

\subsection{UI Representations}

\begin{figure}[t]
\begin{center}
    \centerline{\includegraphics[width=0.9\columnwidth]{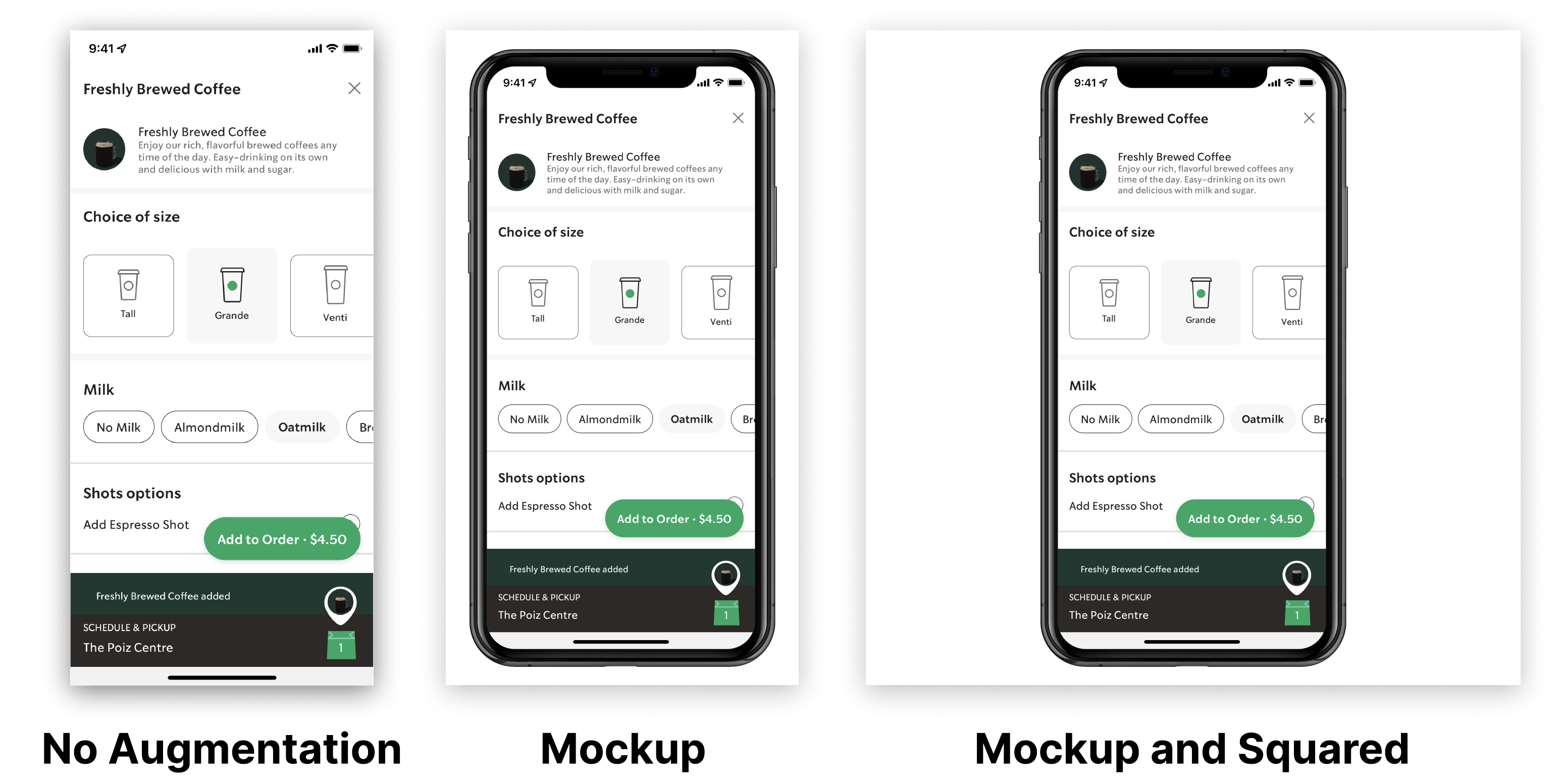}}
    \vspace{-0.3em}
    \caption{Example of our image augmentation for mobile screenshots. We applied the appropriate mockup device according to the resolution and the platform of the screenshot.}
    \label{fig:mockup}
\end{center}
\vspace{-3.3em}
\end{figure}

To ease the use of machine learning models for practitioners, we employed OpenAI's CLIP model \citep{radford2021learning}.
CLIP consists of a visual encoder and a text encoder, both of which are trained using a huge dataset of image-text pairs by optimizing contrastive loss.
So CLIP ensures that the encoded vectors of positive pairs (i.e. aligned image-text pairs) that are closer than those of negative pairs.
By doing so, semantic vectors of both images and texts having similar semantics can be embedded in the vicinity of the joint representation space.
\citet{radford2021learning} further used these encoders on unseen datasets like ImageNet \citep{krizhevsky2017imagenet} for a classification task without any fine-tuning of the model.
It can be done by first embedding all the ImageNet labels and retrieving the closest label for the given image query.
Such application of the model has been named \textit{zero-shot} classification and is a core functionality of foundation models \citep{bommasani2021opportunities}.
We thought that CLIP could understand the representation of UI screenshots, as CLIP understands images that are not seen during training well.

As a sanity check, we first used DALL·E 2 \citep{ramesh2022hierarchical}, which is an image-generation AI that internally uses a noisy version of CLIP, to assess how well CLIP understands the semantics of the UI for the appropriate text prompts associated with the UI screenshots.
By prompting text such as \textit{"A mobile user interface image of shopping application"} to DALL·E 2, we observed that the generated mobile UI images usually come with a mobile device \textit{mockup} rather than just by themselves (\eg, No Augmentation in \cref{fig:mockup}).
Based on this observation, we augmented images with mockups as shown in \cref{fig:mockup} to yield a better representation from CLIP.
We encoded all images with publicly available\footnote{\url{https://github.com/openai/CLIP}} CLIP ViT-L/14\allowbreak@336px%

\paragraph{Uniformity of contrastive representation}
\label{sec:uniformity_method}
Since CLIP embeds images onto the unit hypersphere, we can compute the uniformity loss \citep{wang2020understanding} of an app's screenshots.
Let $I_e \in \mathbb{R}^{n \times d}$ be the set of embedding vectors of UI screenshots that make up an app and let $I_e^{(i)} \in \mathbb{R}^{d}$ be the $i$-th screenshot of the set having a norm of 1.
Then, we can define the uniformity loss \lossu of an app $L_u(I_e)$ as follows: \begin{align}
L_u(I_e) 
& \triangleq \log \frac{1}{n^2} \sum_{i, j} G_t(I_e^{(i)},I_e^{(j)}) \\
& = \log \frac{1}{n^2} \sum_{i, j} e^{2t \cdot I_e^{(i)\mathsf{T}}I_e^{(j)} - 2t}, \quad t > 0, \label{eq:uniformity}
\end{align}
where $G_t$ is a Gaussian potential, and we set $t=2$.

The uniformity loss \lossu ranges from -4 to 0 for $t=2$.
As uniformity loss is negative of uniformity value, lower uniformity loss means the set has more uniformly distributed UI representations.
Since semantically consistent UI screenshots are not uniformly distributed, the lower \lossu also means the lower consistency of the set, and vice versa, the high \lossu for the high consistency of the screenshot set.

\subsection{Optimal Transport}
\label{sec:ot_method}
We employ optimal transport (OT) for app-to-app retrieval, where a transportation plan ${\bf T} \in \mathbb{R}^{n_a \times n_b}_{+}$ is computed to optimize the alignment between two apps $\mathbf{a}$ and $\mathbf{b}$.
We consider apps $\mathbf{a}$ and $\mathbf{b}$ as two discrete distributions $\mathbf{\alpha}$ and $\mathbf{\beta}$, formulated as $\mathbf{\alpha} = \sum_{i=1}^{n_a} \mathbf{n}_i \delta_{\mathbf{a}_i}$ and $\mathbf{\beta} = \sum_{j=1}^{n_b} \mathbf{m}_j \delta_{\mathbf{b}_j}$, where $\mathbf{a}_i$ and $\mathbf{b}_j$ are the embeddings of screenshots making up each app $\mathbf{a}$ and $\mathbf{b}$, and $\delta$ as the Dirac function centered on each screenshot vector.
The marginal plan $\mathbf{n}$ and $\mathbf{m}$ are belong to the $n_a$- and $n_b$-dimensional simplex (i.e., $\sum_{i=1}^{n_a} \mathbf{n}_i=\sum_{j=1}^{n_b} \mathbf{m}_i=1$).
The OT distance between the app $\mathbf{a}$ and $\mathbf{b}$ is then defined as: \begin{align}
    \mathcal{D}_{ot}(\mathbf{a},\mathbf{b}) = \underset{{\bf T} \in \Pi(\mathbf{n},\mathbf{m})}{\text{min}} \sum_{i=1}^{n_a} \sum_{j=1}^{n_b} {\bf T}_{ij} \cdot c(\mathbf{a}_i, \mathbf{b}_j), \label{eq:ot_distance}
\end{align}
where $\Pi(\mathbf{n},\mathbf{m}) = \{ {\bf T} \in \mathbb{R}^{n_a \times n_b}_{+} \vert {\bf T}{\bf 1}_{n_b}=\mathbf{n}, {\bf T}^{\mathsf{T}}{\bf 1}_{n_a}=\mathbf{m} \}$ is a coupling of $\mathbf{n}$ and $\mathbf{m}$, and $c(\cdot, \cdot)$ is the cosine distance.

%% file: Sections/4-Experiments.tex
\subsection{Supremacy of Foundation Models}

To assess how well the foundation model understands the UI semantics, we examined its ability to classify the app category of individual screenshots.
Furthermore, we conducted a Mechanical Turk study, following the settings of \citet{li2021screen2vec},  to compare the retrieval performance of the foundation model against the existing UI semantic representation models.

\label{clip_studies}
\subsubsection{Zero-shot App Category Classification}

\begin{table}[t]
\centering
  \caption{Top-1 and top-5 zero-shot classification accuracies on the Mobbin dataset (19 classes) with our augmentations and prompts.}
  \label{tab:zero_shot_result}
\begin{tabular}{llll}
\toprule
\multicolumn{2}{l}{}                                & Top-1 (↑)         & Top-5 (↑)         \\ \midrule
\multicolumn{2}{c}{Random}                          & 5.26           & 26.31          \\ \midrule
\multirow{3}{*}{\shortstack[l]{ImageNet\\Prompts}} & No augmentation & 32.82          & 65.32          \\
                                  & +Mockup         & 34.33          & 66.10          \\
                                  & +Mockup+Squared & 36.00          & 66.92          \\ \midrule
\multirow{3}{*}{\shortstack[l]{Our\\Prompts}}      & No augmentation & 37.68          & 70.95          \\
                                  & +Mockup         & 39.67          & 74.28          \\
                                  & +Mockup+Squared & \textbf{40.49} & \textbf{74.65} \\ \bottomrule
\end{tabular}
\vspace{-1.5em}
\end{table}

As mentioned earlier, the foundation model we used (CLIP) has the capability to perform image classification in a zero-shot manner without additional training.
The zero-shot classification proceeds in the following order: first, we prepare appropriate text prompts those match each class (\eg, A photo of a \{class\}).
Second, we extract text features for each prompt using the CLIP text encoder.
Lastly, the top-k labels are predicted by the cosine similarity between their text features and the query image feature.

Each screenshot in the Mobbin dataset is classified by CLIP according to its app category shown in \cref{fig:dataset_chart}.
Furthermore, to demonstrate the effectiveness of data augmentation in \cref{fig:mockup}, we compared the classification accuracy between the original screenshots, the screenshots added mockup template, and the square-resized version of them.
There are a total of 19 app categories in the Mobbin dataset, thus it has 5.2 and 26.3 by-chance accuracy for top-1 and top-5 classification, respectively.
Original CLIP paper uses seven text prompts \footnote{\url{https://github.com/openai/CLIP/blob/main/notebooks/Prompt_Engineering_for_ImageNet.ipynb}} (\eg, \textit{itap of a \{category\}., a photo of the [small/large] \{category\}}) for ImageNet zero-shot classification.
Since the prompts are engineered for ImageNet's images, which are mainly made up of photos of an object, we redesigned twelve prompts for UI screenshots, each of which includes \textit{user interfaces}, \textit{UI}, \textit{mobile screen}, or \textit{screenshot} (\eg, \textit{a screenshot of \{category\} app}, \textit{A user interface of \{category\} application.}, \textit{UI of \{category\} app.})

\cref{tab:zero_shot_result} shows overall zero-shot app category classification results on the Mobbin dataset.
Based on each augmentation result, we can confirm that the CLIP's understanding of UI semantics was improved when our augmentations, mockup templates, and square-resizing are applied.
The same classification, but with the custom text prompts designed for the UI screenshots resulted in superior performance when compared to those of prompts for the natural photos (ImageNet).
CLIP showed top-1 accuracy of approximately six times (with naive CLIP setting) to eight times (with our engineering) better than randomly classifying the original screenshots without additional training.
This result indicates that CLIP has a certain level of understanding of user interfaces and application categories.

\subsubsection{Screenshot Retrieval Comparison}
\label{mturk_study}

To find out whether the foundation model's retrieval results are preferred by humans or not, we conducted a comparative Mechanical Turk user study proposed by \citet{li2021screen2vec} and compared the results with Screen2Vec \citep{li2021screen2vec}.
To carry out the experiment fairly, we used the Rico dataset for the retrieval experiment following the settings and baselines in~\citet{li2021screen2vec}. %
The models used for their Mechanical Turk experiments are as follows:
\paragraph{Screen2Vec}
Screen2Vec uses Rico's screenshot and its metadata.
The GUI components and their location information are encoded into a 768-dimensional vector.
Plus, the application information is extracted using Sentence-BERT resulting in another 768-dimensional vector, then the two vectors are concatenated into a 1536-dimensional GUI screen embedding vector.
\paragraph{TextOnly}
This variant reproduces the models proposed in \citet{li2020multi}.
It extracts text features of all texts in a screenshot using pre-trained Sentence-BERT \citep{reimers2019sentence} and averages the features into a single 768-dimensional vector.
\paragraph{LayoutOnly}
This variant reproduces the autoencoder model proposed in \citet{deka2017rico}.
It embeds the layout of a screenshot into a 64-dimensional feature.

CLIP image features were extracted both with mockup templates and square-resized augmentation to quantitatively prove the effect of image augmentation in retrieval tasks.
We randomly selected 50 screenshots from Enrico \citep{leiva2020enrico}, which is a curated subset of Rico.
Using these screenshots as the queries, the top-5 screenshots were retrieved from the entire Rico dataset for each model.
Since each of the five models (\ie, CLIP, CLIP+aug, Screen2Vec, TextOnly, LayoutOnly) retrieves five screenshots, a total of 25 similar images were obtained for each query screenshot.
We removed duplicates from the retrieved, thus making fewer than 25 screenshots for some of the queries, resulting in a total of 999 pairs of screens (query and retrieved screen), averaging 19.98 per query screen.

Screen pair similarity was measured using the criteria from \citet{li2021screen2vec}, including app similarity (likeness between two apps), screen type similarity (parity in the roles of two screens), and content similarity (congruity of the displayed contents).
Crowd workers were asked to measure the similarity of five sets of query screenshots on a five-point Likert scale and the corresponding retrieved screenshots for each task, and five different workers measured similarities for each pair of screenshots.
We paid two dollars for each batch of five source screens and it took an average of 28 minutes.

\begin{table}[t]
\centering
  \caption{The mean similarity score and its standard deviation ($N=1250$ each) were measured by Mechanical Turk workers. A higher score means more preferred retrieval results. The Mann-Whitney U test shows our models are statistically significantly better than Screen2Vec and other models ($p < 0.0001$).}
  \label{tab:mturk_result}
\begin{tabular}{lll}
\toprule
\multicolumn{2}{l}{}                    & Score (↑) \\ \midrule
\citet{li2021screen2vec}  & Screen2Vec & 2.92±1.36 \\
\citet{li2020multi} & TextOnly & 3.22±1.30 \\
\citet{deka2017rico} & LayoutOnly & 3.00±1.33 \\ \midrule
\multirow{2}{*}{Ours} & No augmentation & 3.50±1.16 \\
                      & +Mockup+Squared & \textbf{3.57±1.08} \\ \bottomrule
\end{tabular}
\end{table}

\cref{tab:mturk_result} reports the mean similarity value measured by the crowd worker for each model.
The study revealed a worker preference for screenshots retrieved via CLIP image features, outperforming the three comparative models (Screen2Vec, TextOnly, LayoutOnly).
Moreover, the utilization of mockup templates and square-resize augmentation enhanced the assessment of CLIP.
Notably, the Mann-Whitney U test underscored the superior performance of the two CLIP models with statistical significance ($p < 0.0001$).

\subsection{App-to-App Retrieval}

\begin{figure}[t]
    \centering
    \includegraphics[width=\linewidth]{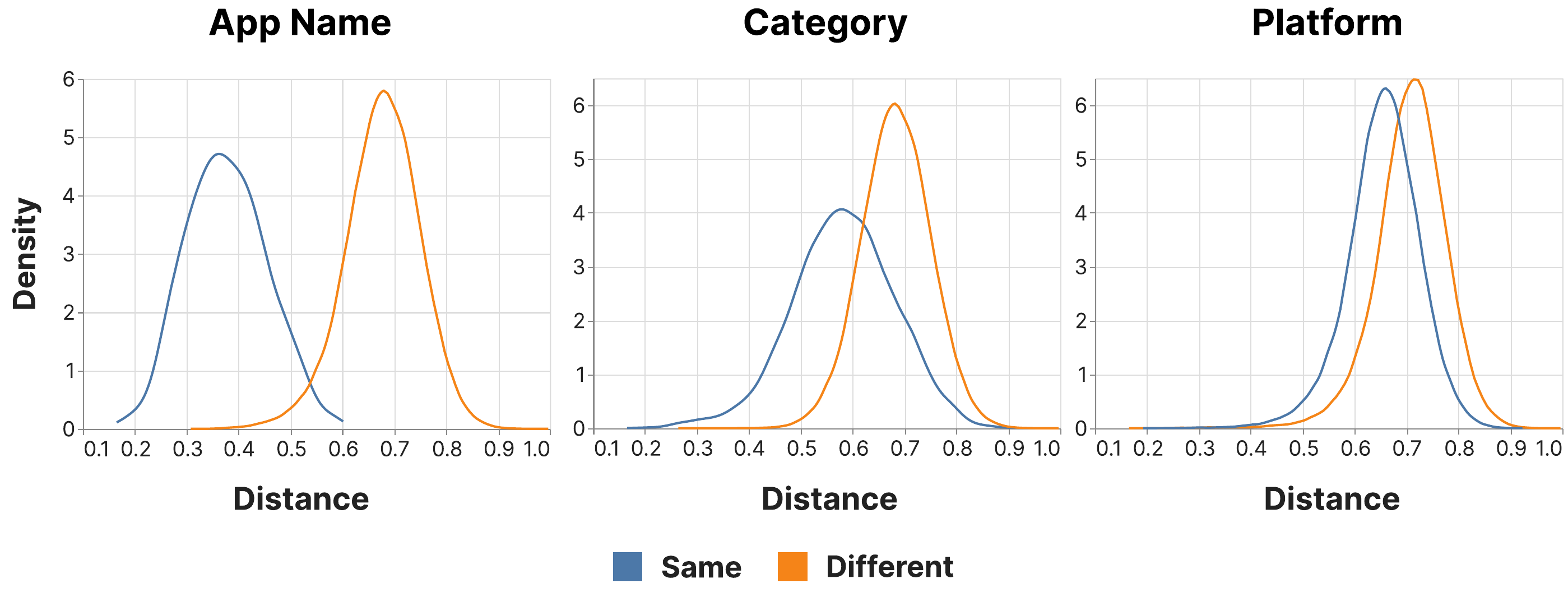}
    \vspace{-1.5em}
    \caption{Distribution of pairwise distance of applications on three criteria: (left) app name, (middle) app category, and (right) platform (operating system).}
    \label{fig:pair_distance}
    \vspace{-1.8em}

\end{figure}

\begin{figure}[t]
    \centering
    \includegraphics[width=\linewidth]{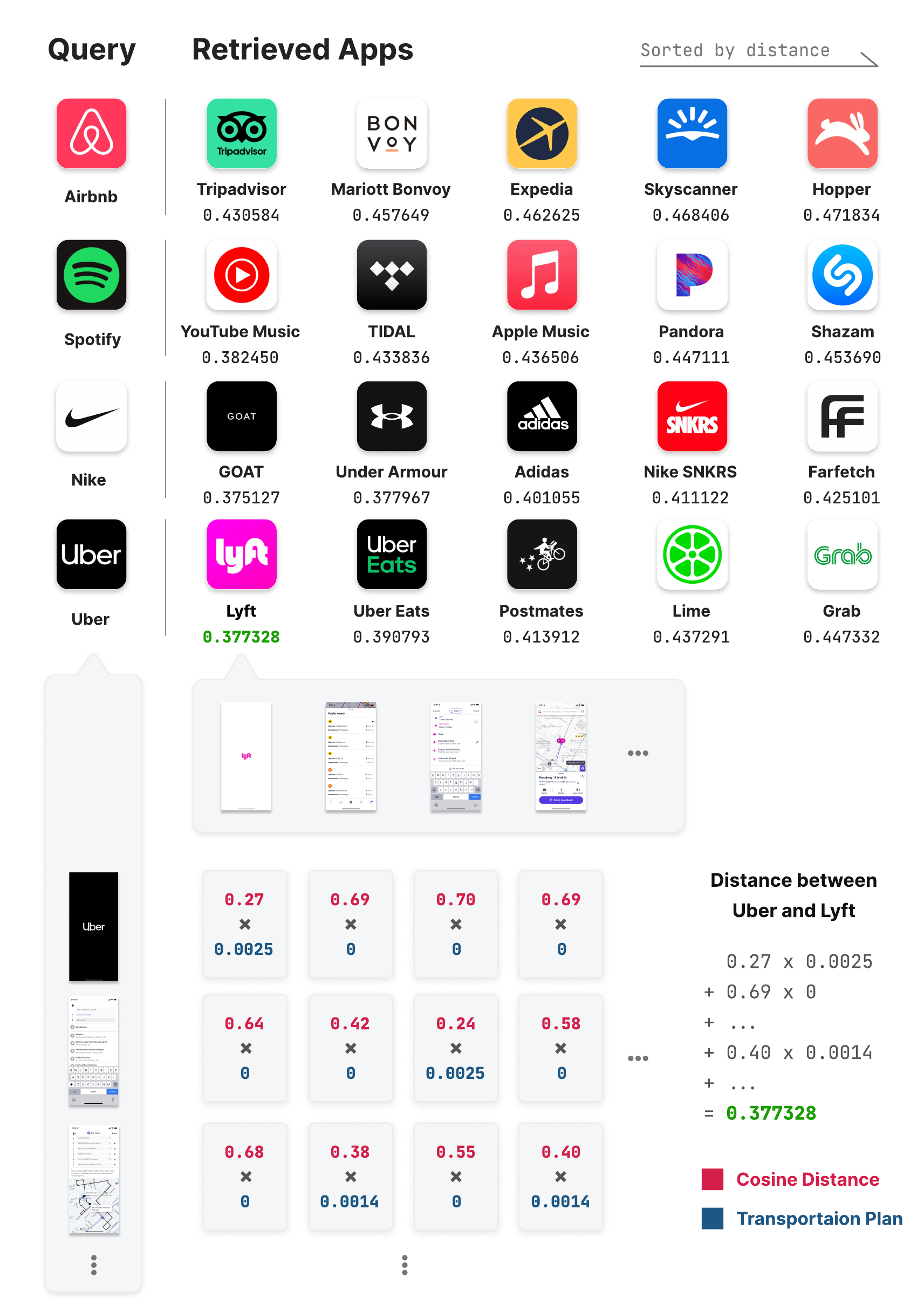}
    \vspace{-1.5em}
    \caption{(Left) The list of retrieved apps for given query apps across various categories; retrieved apps are sorted with their \dot to their query. (Right) The detailed process of getting \dot, as described in \cref{sec:ot_method}.}
    \label{fig:app_retrieval}
    \vspace{-1.5em}

\end{figure}

To extend data-driven UI inspiration beyond simple single screenshot-level retrieval, we introduce the concept of optimal transport to properly implement \appt retrieval.
\cref{fig:app_retrieval} illustrates the detailed process of app-to-app retrieval using optimal transport by example.
First, we extract the semantic representation of the augmented screenshot in each application using the CLIP encoder we used throughout the paper.
Using these features, we obtain the pairwise cosine distance matrix between the UI semantic vector sets of two applications.
We assign uniform mass over the screenshots to form a uniform distribution, which makes the initial $n$ and $m$ in \cref{sec:ot_method} to a uniform distribution.
Then using the pairwise distance matrix and the distributions, we solve the optimal transport problem using POT \citep{flamary2021pot} as described in \cref{sec:ot_method} to get an optimal transportation plan.
Finally, we calculate a 1-Wasserstein distance (\ie, \dot) between the application pair by element-wise multiplying the distance matrix with the transportation plan matrix and summing the elements up, following \cref{eq:ot_distance}.
To assess \appt retrieval based on \dot, we perform a quantitative analysis of \dot as a distance between applications and conducted a case study on the possible applications of the transportation matrix.

\paragraph{\dot as an app-to-app distance}
We calculated \dot for 306,153 pairs of apps for a total of 783 apps of 319 unique app names\footnote{We use the term \textit{name} to indicate a universally unique identifier (UUID); thus, different apps with the same name are treated individually in this paper.} in the Mobbin dataset.
The calculation took about 32 minutes to process all pairs, which is about 158 app pairs per second using a machine with a single NVIDIA Titan RTX GPU.
Using \dot of the pairs, we analyze the statistics of the Mobbin dataset in three criteria: each app's name, category, and platform (iOS or Android).
\cref{fig:pair_distance} shows distance distribution by the app's name, category, and platform, respectively.
There are 944 pairs sharing an app name but differing in version (platform or release date), 21,141 pairs with identical categories, and 180,603 pairs on the same platform
Apps with the same app name, category, and platform are displayed in blue, whereas apps with a different app name, category, and platform are displayed in orange.
Across all three cases, the distance was significantly shorter when the app pairs shared the same app name, as the overall composition and semantics of screenshots in the app do not change much for different releases of the app.
This outcome aligns with the designers' intent of maintaining app identity through updates or cross-platform deployment, affirming the suitability of \dot for modeling app distance
While category and platform act as group identifiers and are thus less unique than an app's name, our \dot model effectively demonstrated a shorter \dot for apps sharing the same category/platform compared to those differing in these criteria.
Notably, \dot also revealed that platform information, representing a larger group of apps, is more general than category information, evidenced by a smaller \dot distribution gap for the platform criterion.
We want to note that it is a very significant result since the figure is drawn with more than 300,000 pairs of pairs.

\paragraph{Interpretability of an optimal transportation plan}

Besides the dataset-wide quantitative analysis, we highlight a few examples of retrieved apps for given queries in \cref{fig:app_retrieval}.
The figure demonstrates the method's efficacy by showcasing retrieved apps, from various categories like Airbnb, Spotify, Nike, and Uber, that bear similar semantics.
Impressively, these results were acquired using merely app screenshots, sans any metadata like app category or component hierarchy.
The transport plan, or optimal transport matrix, describes how to optimally move masses when there are two distributions.
The transport plan, or optimal transport matrix, describes the optimal mass movement between two distributions, enabling the identification of similar screenshots between two apps, as they will exhibit similar vector representations.

\subsection{Design Consistency Check}

\begin{figure}[t]
    \centering
    \includegraphics[width=\linewidth]{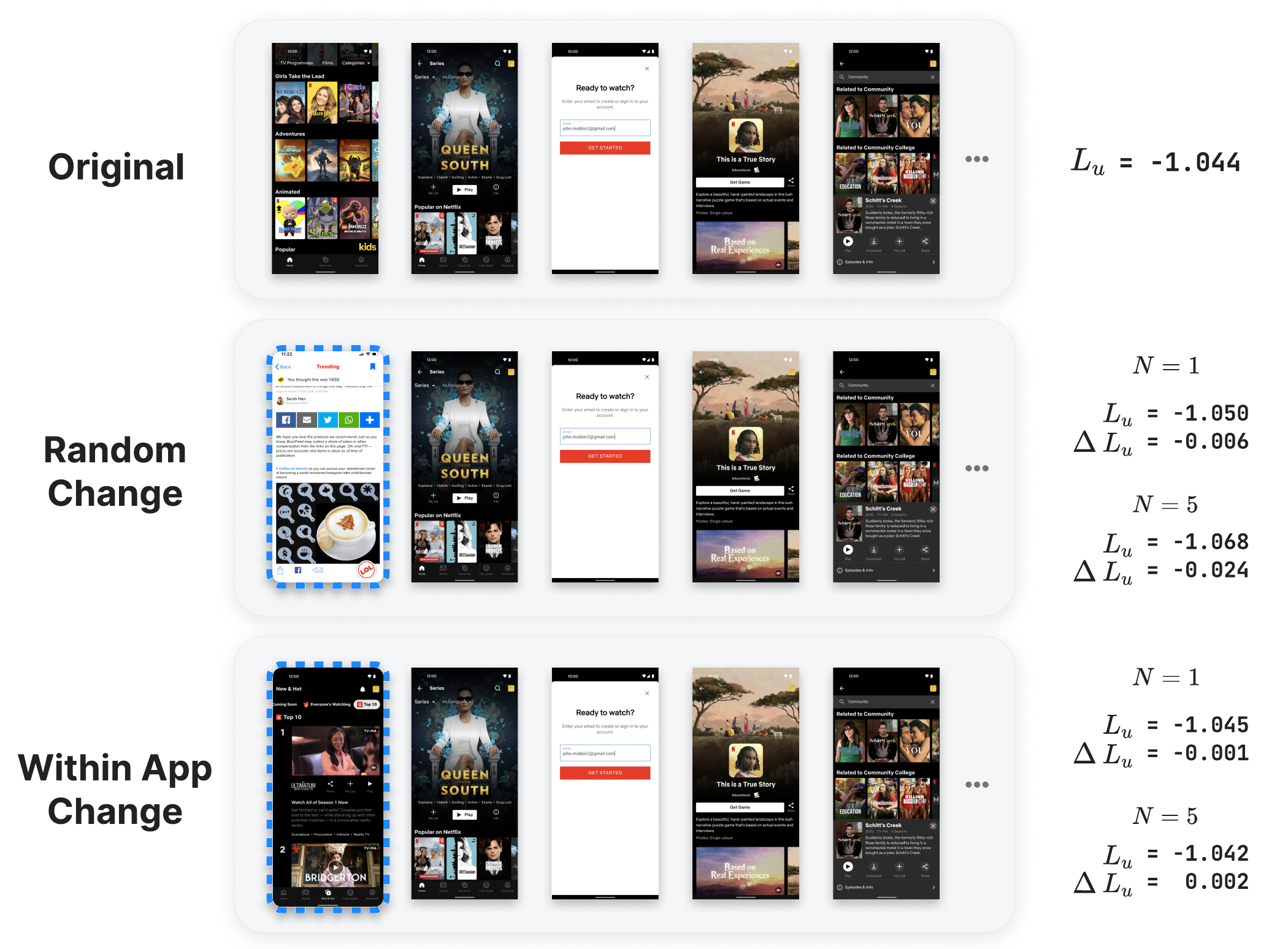}
    \vspace{-1.5em}
    \caption{Our experiment conditions and sample result (Netflix app) of validating uniformity loss \deltalu as design consistency check metric.}
    \label{fig:uniformity_images}
\end{figure}

\begin{figure}[t]
    \centering
    \includegraphics[width=\linewidth]{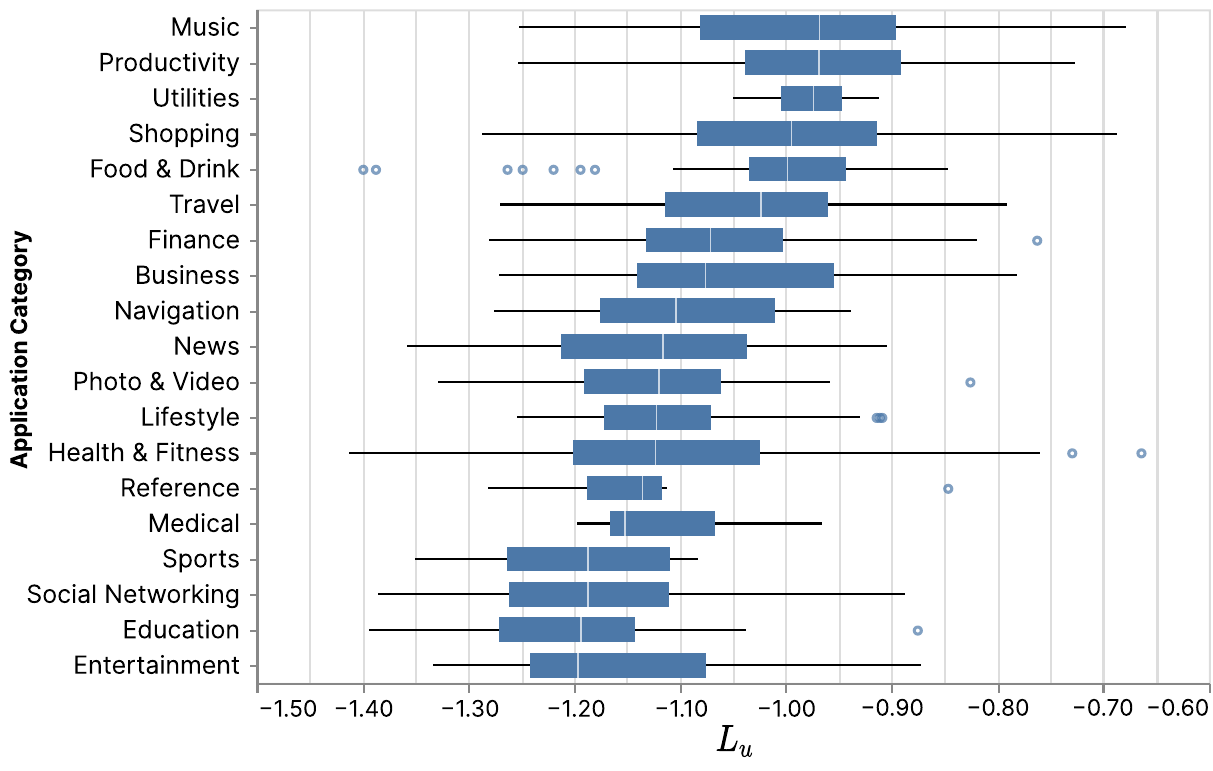}
    \vspace{-1.5em}
    \caption{Distribution of \lossu by app category, sorted by median}
    \label{fig:uniformity_category}
    \vspace{-1.5em}
\end{figure}

\begin{figure}[t]
    \centering
    \includegraphics[width=\linewidth]{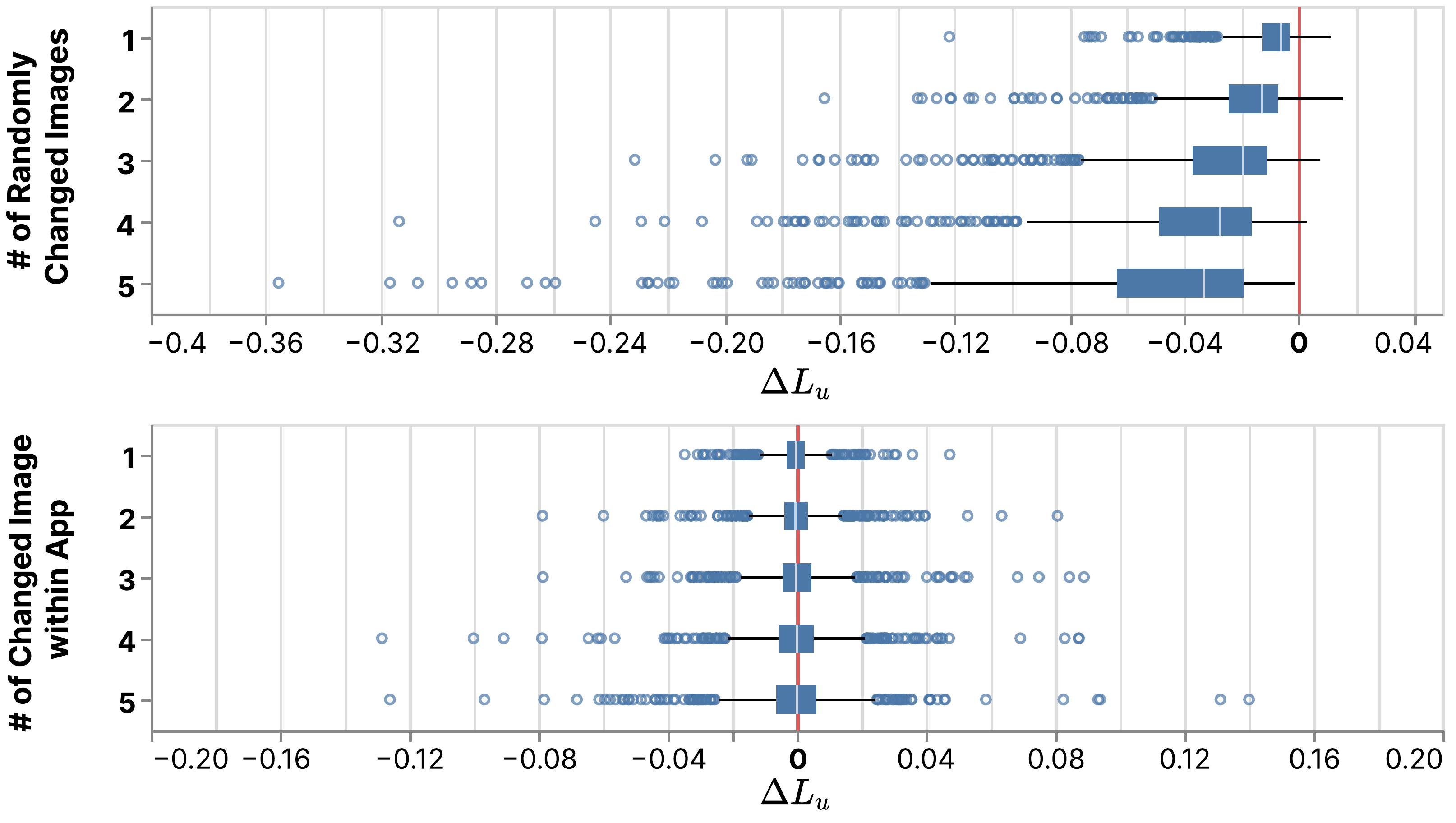}
    \vspace{-1.5em}
    \caption{The amount of change of uniformity loss \deltalu according to our experimental conditions. (Top) \deltalu when we change the images in the app randomly, (Bottom) \deltalu when we change the images in the app with the reserved screenshots from the same app}
    \label{fig:uniformity}
    \vspace{-1.5em}

\end{figure}

As described in \cref{sec:uniformity_method}, we use \lossu to measure the consistency of an app. 
We calculated \lossu for every app in the dataset, \cref{fig:uniformity_category} shows statistics of \lossu grouped by the app categories.
\lossu in the dataset ranges from $-1.41$ to $-0.66$.
Categories such as \textit{entertainment, education, and social networking} turned out to have a lower \lossu, indicative of inconsistent and diverse screenshots, owing to their UI screens frequently consisting of various media types.
On the other hand, the latter categories generally consist of their UI with icons and symbols rather than media, thus resulting in high \lossu.

To test whether \lossu could serve as a metric for data-driven design consistency check discussed in \cref{sec:uniformity_background}, we designed two studies on the Mobbin dataset.
The first, simulating a hypothetical scenario, involved designers assessing the alignment of new UI screen drafts with existing screens.
This process, labeled as \textit{Random Change} in \cref{fig:uniformity_images}, consisted of substituting $N$ images per app with random ones from other apps in the dataset and subsequently calculating \lossu.
Testing different $N$ values facilitated an examination of the robustness of our uniformity metric in relation to the set size intended for inspection by the designer.

The difference between two $L_u$s (\deltalu) is shown in the first figure of \cref{fig:uniformity}.
As shown in the figure, \deltalu decreases as the number of randomly changed images increases.
Through the t-test, the drop of \lossu is statistically significant regardless of $N$ ($p < 0.0001$).

The second study is to test whether this drop occurs within the same design semantics, as the metric would be meaningless if it drops regardless of the semantics of the changed images.
To implement the study, we first set aside five UI screenshot images from each app\footnote{For brevity, we also used this setting that set aside five UI screenshots while measuring \lossu of the random screenshot change experiment.} in the dataset.
Subsequently, we replaced $N$ images for each app with the images we had set aside, as opposed to images from random apps as in the previous study.
The second figure of \cref{fig:uniformity} indicates no changes in \deltalu in this scenario.
A t-test confirmed that \lossu remained constant irrespective of $N$ ($p > 0.28$).
These studies collectively underscore \lossu as a robust proxy for assessing design consistency.

%% file: Sections/5-Limitation.tex
Although we explored the conceptual applications of app-to-app retrieval and in-app design consistency based on various reports on designers' preferences \citep{colusso2017translational,wu2021exploring,lu2022bridging}, it is still required to prove their efficacy in the real environment.
As such studies require conducting formative studies and in-depth user studies, making it out of the scope of our paper proposing novel applications of UI representation models, we leave it for future work.

Optimal transport allows various initial marginal distributions, assigning more \textit{mass} to one screenshot than another.
Since we assumed the condition with no information on which screenshot is more important than another in the app, we used a uniform distribution throughout the paper.
A uniform distribution is a good choice to admit the maximum entropy probability distribution \citep{bernardo2009bayesian}, but assigning different initial marginal distributions could enable different applications such as focusing more on certain screenshots or neglecting user-selected UI components during \dot computation.
The manipulation of the initial marginal distribution remains for future work.

Another possible future work could be to extend the dataset to group screenshots with their feature or functionality rather than an app.
Although we only tested primitive app groups of screenshots, since an app is made up of at least a few dozen screenshots, grouping screenshots with other criteria would be much easier to make a dataset since it requires fewer screenshots per group and yields different retrieval opportunities.

Regarding design consistency, \lossu is innately a relative metric, and hence, cannot evaluate design quality in absolute terms.
Although it was not our focus to build absolute guidelines, future work could involve defining a golden standard for specific design principles and utilizing \deltalu to quantify the deviation of the query set from this standard.

%% file: Sections/6-Conclusion.tex
In this paper, we envisioned the applications of UI representation models including app-to-app retrieval and design consistency check.
As the first stage of this research, we investigated the supremacy of zero-shot adaptation of a foundation model, CLIP, and how its representation appeals to humans by conducting a Mechanical Turk study on a single-screenshot retrieval task.
Using the CLIP UI representation, we devised (1) \dot that measures the distance between apps by computing the optimal transportation plan between their UI screenshots, and (2) \lossu that measures the semantic design consistency of an app by computing the pairwise Gaussian potentials between the UI screenshots of the app.
Through multiple proofs of concept and analysis on our newly collected Mobbin dataset, we showed that both \dot and \lossu are valid metrics for app-to-app retrieval and in-app design consistency check, respectively.
We would like to highlight that our proposed methods can be executed on a personal laptop without expensive equipment such as GPU clusters while opening numerous opportunities for computational UI engineering. Going forward, we are excited to continue our endeavors toward building interfaces for designers that are equipped with our computational approaches.

%% file: example_paper.bbl
\begin{thebibliography}{33}
\providecommand{\natexlab}[1]{#1}
\providecommand{\url}[1]{\texttt{#1}}
\expandafter\ifx\csname urlstyle\endcsname\relax
  \providecommand{\doi}[1]{doi: #1}\else
  \providecommand{\doi}{doi: \begingroup \urlstyle{rm}\Url}\fi

\bibitem[Bernardo \& Smith(2009)Bernardo and Smith]{bernardo2009bayesian}
Bernardo, J.~M. and Smith, A.~F.
\newblock \emph{Bayesian theory}, volume 405.
\newblock John Wiley \& Sons, 2009.

\bibitem[Bommasani et~al.(2021)Bommasani, Hudson, Adeli, Altman, Arora, von
  Arx, Bernstein, Bohg, Bosselut, Brunskill,
  et~al.]{bommasani2021opportunities}
Bommasani, R., Hudson, D.~A., Adeli, E., Altman, R., Arora, S., von Arx, S.,
  Bernstein, M.~S., Bohg, J., Bosselut, A., Brunskill, E., et~al.
\newblock On the opportunities and risks of foundation models.
\newblock \emph{arXiv preprint arXiv:2108.07258}, 2021.

\bibitem[Bonnardel(1999)]{bonnardel1999creativity}
Bonnardel, N.
\newblock Creativity in design activities: The role of analogies in a
  constrained cognitive environment.
\newblock In \emph{Proceedings of the 3rd conference on Creativity \&
  cognition}, pp.\  158--165, 1999.

\bibitem[Brown et~al.(2020)Brown, Mann, Ryder, Subbiah, Kaplan, Dhariwal,
  Neelakantan, Shyam, Sastry, Askell, et~al.]{brown2020language}
Brown, T., Mann, B., Ryder, N., Subbiah, M., Kaplan, J.~D., Dhariwal, P.,
  Neelakantan, A., Shyam, P., Sastry, G., Askell, A., et~al.
\newblock Language models are few-shot learners.
\newblock \emph{Advances in neural information processing systems},
  33:\penalty0 1877--1901, 2020.

\bibitem[Bunian et~al.(2021)Bunian, Li, Jemmali, Harteveld, Fu, and Seif
  El-Nasr]{bunian2021vins}
Bunian, S., Li, K., Jemmali, C., Harteveld, C., Fu, Y., and Seif El-Nasr, M.~S.
\newblock Vins: Visual search for mobile user interface design.
\newblock In \emph{Proceedings of the 2021 CHI Conference on Human Factors in
  Computing Systems}, CHI '21, New York, NY, USA, 2021. Association for
  Computing Machinery.
\newblock ISBN 9781450380966.
\newblock \doi{10.1145/3411764.3445762}.
\newblock URL \url{https://doi.org/10.1145/3411764.3445762}.

\bibitem[Burny \& Vanderdonckt(2022)Burny and Vanderdonckt]{burny2022semi}
Burny, N. and Vanderdonckt, J.
\newblock (semi-) automatic computation of user interface consistency.
\newblock In \emph{Companion of the 2022 ACM SIGCHI Symposium on Engineering
  Interactive Computing Systems}, pp.\  5--13, 2022.

\bibitem[Chen et~al.(2020)Chen, Chen, Xing, Xia, Zhu, Grundy, and
  Wang]{chen2020wireframe}
Chen, J., Chen, C., Xing, Z., Xia, X., Zhu, L., Grundy, J., and Wang, J.
\newblock Wireframe-based ui design search through image autoencoder.
\newblock \emph{ACM Transactions on Software Engineering and Methodology
  (TOSEM)}, 29\penalty0 (3):\penalty0 1--31, 2020.

\bibitem[Colusso et~al.(2017)Colusso, Bennett, Hsieh, and
  Munson]{colusso2017translational}
Colusso, L., Bennett, C.~L., Hsieh, G., and Munson, S.~A.
\newblock Translational resources: Reducing the gap between academic research
  and hci practice.
\newblock In \emph{Proceedings of the 2017 Conference on Designing Interactive
  Systems}, pp.\  957--968, 2017.

\bibitem[Deka et~al.(2017)Deka, Huang, Franzen, Hibschman, Afergan, Li,
  Nichols, and Kumar]{deka2017rico}
Deka, B., Huang, Z., Franzen, C., Hibschman, J., Afergan, D., Li, Y., Nichols,
  J., and Kumar, R.
\newblock Rico: A mobile app dataset for building data-driven design
  applications.
\newblock In \emph{Proceedings of the 30th Annual ACM Symposium on User
  Interface Software and Technology}, UIST '17, pp.\  845–854, New York, NY,
  USA, 2017. Association for Computing Machinery.
\newblock ISBN 9781450349819.
\newblock \doi{10.1145/3126594.3126651}.
\newblock URL \url{https://doi.org/10.1145/3126594.3126651}.

\bibitem[Flamary et~al.(2021)Flamary, Courty, Gramfort, Alaya, Boisbunon,
  Chambon, Chapel, Corenflos, Fatras, Fournier, Gautheron, Gayraud, Janati,
  Rakotomamonjy, Redko, Rolet, Schutz, Seguy, Sutherland, Tavenard, Tong, and
  Vayer]{flamary2021pot}
Flamary, R., Courty, N., Gramfort, A., Alaya, M.~Z., Boisbunon, A., Chambon,
  S., Chapel, L., Corenflos, A., Fatras, K., Fournier, N., Gautheron, L.,
  Gayraud, N.~T., Janati, H., Rakotomamonjy, A., Redko, I., Rolet, A., Schutz,
  A., Seguy, V., Sutherland, D.~J., Tavenard, R., Tong, A., and Vayer, T.
\newblock Pot: Python optimal transport.
\newblock \emph{Journal of Machine Learning Research}, 22\penalty0
  (78):\penalty0 1--8, 2021.
\newblock URL \url{http://jmlr.org/papers/v22/20-451.html}.

\bibitem[Herring et~al.(2009)Herring, Chang, Krantzler, and
  Bailey]{herring2009getting}
Herring, S.~R., Chang, C.-C., Krantzler, J., and Bailey, B.~P.
\newblock Getting inspired! understanding how and why examples are used in
  creative design practice.
\newblock In \emph{Proceedings of the SIGCHI Conference on Human Factors in
  Computing Systems}, CHI '09, pp.\  87–96, New York, NY, USA, 2009.
  Association for Computing Machinery.
\newblock ISBN 9781605582467.
\newblock \doi{10.1145/1518701.1518717}.
\newblock URL \url{https://doi.org/10.1145/1518701.1518717}.

\bibitem[Huang et~al.(2019)Huang, Canny, and Nichols]{huang2019swire}
Huang, F., Canny, J.~F., and Nichols, J.
\newblock Swire: Sketch-based user interface retrieval.
\newblock In \emph{Proceedings of the 2019 CHI Conference on Human Factors in
  Computing Systems}, CHI '19, pp.\  1–10, New York, NY, USA, 2019.
  Association for Computing Machinery.
\newblock ISBN 9781450359702.
\newblock \doi{10.1145/3290605.3300334}.
\newblock URL \url{https://doi.org/10.1145/3290605.3300334}.

\bibitem[Ivory \& Hearst(2001)Ivory and Hearst]{ivory2001state}
Ivory, M.~Y. and Hearst, M.~A.
\newblock The state of the art in automating usability evaluation of user
  interfaces.
\newblock \emph{ACM Computing Surveys (CSUR)}, 33\penalty0 (4):\penalty0
  470--516, 2001.

\bibitem[Jiang et~al.(2022)Jiang, Lu, Nichols, Stuerzlinger, Yu, Lutteroth, Li,
  Kumar, and Li]{jiang2022computational}
Jiang, Y., Lu, Y., Nichols, J., Stuerzlinger, W., Yu, C., Lutteroth, C., Li,
  Y., Kumar, R., and Li, T. J.-J.
\newblock Computational approaches for understanding, generating, and adapting
  user interfaces.
\newblock In \emph{CHI Conference on Human Factors in Computing Systems
  Extended Abstracts}, pp.\  1--6, 2022.

\bibitem[Krizhevsky et~al.(2017)Krizhevsky, Sutskever, and
  Hinton]{krizhevsky2017imagenet}
Krizhevsky, A., Sutskever, I., and Hinton, G.~E.
\newblock Imagenet classification with deep convolutional neural networks.
\newblock \emph{Communications of the ACM}, 60\penalty0 (6):\penalty0 84--90,
  2017.

\bibitem[Kumar et~al.(2013)Kumar, Satyanarayan, Torres, Lim, Ahmad, Klemmer,
  and Talton]{kumar2013webzeitgeist}
Kumar, R., Satyanarayan, A., Torres, C., Lim, M., Ahmad, S., Klemmer, S.~R.,
  and Talton, J.~O.
\newblock Webzeitgeist: design mining the web.
\newblock In \emph{Proceedings of the SIGCHI Conference on Human Factors in
  Computing Systems}, pp.\  3083--3092, 2013.

\bibitem[Lee et~al.(2010)Lee, Srivastava, Kumar, Brafman, and
  Klemmer]{lee2010designing}
Lee, B., Srivastava, S., Kumar, R., Brafman, R., and Klemmer, S.~R.
\newblock Designing with interactive example galleries.
\newblock In \emph{Proceedings of the SIGCHI Conference on Human Factors in
  Computing Systems}, CHI '10, pp.\  2257–2266, New York, NY, USA, 2010.
  Association for Computing Machinery.
\newblock ISBN 9781605589299.
\newblock \doi{10.1145/1753326.1753667}.
\newblock URL \url{https://doi.org/10.1145/1753326.1753667}.

\bibitem[Leiva et~al.(2020)Leiva, Hota, and Oulasvirta]{leiva2020enrico}
Leiva, L.~A., Hota, A., and Oulasvirta, A.
\newblock Enrico: A dataset for topic modeling of mobile ui designs.
\newblock In \emph{22nd International Conference on Human-Computer Interaction
  with Mobile Devices and Services}, pp.\  1--4, 2020.

\bibitem[Li et~al.(2020)Li, Chen, Xia, Mitchell, and Myers]{li2020multi}
Li, T. J.-J., Chen, J., Xia, H., Mitchell, T.~M., and Myers, B.~A.
\newblock Multi-modal repairs of conversational breakdowns in task-oriented
  dialogs.
\newblock In \emph{Proceedings of the 33rd Annual ACM Symposium on User
  Interface Software and Technology}, pp.\  1094--1107, 2020.

\bibitem[Li et~al.(2021)Li, Popowski, Mitchell, and Myers]{li2021screen2vec}
Li, T. J.-J., Popowski, L., Mitchell, T., and Myers, B.~A.
\newblock Screen2vec: Semantic embedding of gui screens and gui components.
\newblock In \emph{Proceedings of the 2021 CHI Conference on Human Factors in
  Computing Systems}, CHI '21, New York, NY, USA, 2021. Association for
  Computing Machinery.
\newblock ISBN 9781450380966.
\newblock \doi{10.1145/3411764.3445049}.
\newblock URL \url{https://doi.org/10.1145/3411764.3445049}.

\bibitem[Liu et~al.(2018)Liu, Craft, Situ, Yumer, Mech, and
  Kumar]{liu2018learning}
Liu, T.~F., Craft, M., Situ, J., Yumer, E., Mech, R., and Kumar, R.
\newblock Learning design semantics for mobile apps.
\newblock In \emph{Proceedings of the 31st Annual ACM Symposium on User
  Interface Software and Technology}, UIST '18, pp.\  569–579, New York, NY,
  USA, 2018. Association for Computing Machinery.
\newblock ISBN 9781450359481.
\newblock \doi{10.1145/3242587.3242650}.
\newblock URL \url{https://doi.org/10.1145/3242587.3242650}.

\bibitem[Lu et~al.(2022)Lu, Zhang, Zhang, and Li]{lu2022bridging}
Lu, Y., Zhang, C., Zhang, I., and Li, T. J.-J.
\newblock Bridging the gap between ux practitioners’ work practices and
  ai-enabled design support tools.
\newblock In \emph{CHI Conference on Human Factors in Computing Systems
  Extended Abstracts}, pp.\  1--7, 2022.

\bibitem[Mahajan \& Shneiderman(1997)Mahajan and
  Shneiderman]{mahajan1997visual}
Mahajan, R. and Shneiderman, B.
\newblock Visual and textual consistency checking tools for graphical user
  interfaces.
\newblock \emph{IEEE Transactions on software engineering}, 23\penalty0
  (11):\penalty0 722--735, 1997.

\bibitem[Peyr{\'e} et~al.(2019)Peyr{\'e}, Cuturi,
  et~al.]{peyre2019computational}
Peyr{\'e}, G., Cuturi, M., et~al.
\newblock Computational optimal transport: With applications to data science.
\newblock \emph{Foundations and Trends{\textregistered} in Machine Learning},
  11\penalty0 (5-6):\penalty0 355--607, 2019.

\bibitem[Radford et~al.(2021)Radford, Kim, Hallacy, Ramesh, Goh, Agarwal,
  Sastry, Askell, Mishkin, Clark, et~al.]{radford2021learning}
Radford, A., Kim, J.~W., Hallacy, C., Ramesh, A., Goh, G., Agarwal, S., Sastry,
  G., Askell, A., Mishkin, P., Clark, J., et~al.
\newblock Learning transferable visual models from natural language
  supervision.
\newblock In \emph{International Conference on Machine Learning}, pp.\
  8748--8763. PMLR, 2021.

\bibitem[Ramesh et~al.(2022)Ramesh, Dhariwal, Nichol, Chu, and
  Chen]{ramesh2022hierarchical}
Ramesh, A., Dhariwal, P., Nichol, A., Chu, C., and Chen, M.
\newblock Hierarchical text-conditional image generation with clip latents.
\newblock \emph{arXiv preprint arXiv:2204.06125}, 2022.

\bibitem[Reimers \& Gurevych(2019)Reimers and Gurevych]{reimers2019sentence}
Reimers, N. and Gurevych, I.
\newblock Sentence-bert: Sentence embeddings using siamese bert-networks.
\newblock \emph{arXiv preprint arXiv:1908.10084}, 2019.

\bibitem[Ritchie et~al.(2011)Ritchie, Kejriwal, and Klemmer]{ritchie2011d}
Ritchie, D., Kejriwal, A.~A., and Klemmer, S.~R.
\newblock d. tour: Style-based exploration of design example galleries.
\newblock In \emph{Proceedings of the 24th annual ACM symposium on User
  interface software and technology}, pp.\  165--174, 2011.

\bibitem[Santambrogio(2015)]{santambrogio2015optimal}
Santambrogio, F.
\newblock Optimal transport for applied mathematicians.
\newblock \emph{Birk{\"a}user, NY}, 55\penalty0 (58-63):\penalty0 94, 2015.

\bibitem[Villani(2009)]{villani2009optimal}
Villani, C.
\newblock \emph{Optimal transport: old and new}, volume 338.
\newblock Springer, 2009.

\bibitem[Wang \& Isola(2020)Wang and Isola]{wang2020understanding}
Wang, T. and Isola, P.
\newblock Understanding contrastive representation learning through alignment
  and uniformity on the hypersphere.
\newblock In \emph{International Conference on Machine Learning}, pp.\
  9929--9939. PMLR, 2020.

\bibitem[Wu et~al.(2021)Wu, Xu, Liu, Peng, Xu, and Ma]{wu2021exploring}
Wu, Z., Xu, Q., Liu, Y., Peng, Z., Xu, Y., and Ma, X.
\newblock Exploring designers’ practice of online example management for
  supporting mobile ui design.
\newblock In \emph{Proceedings of the 23rd International Conference on Mobile
  Human-Computer Interaction}, pp.\  1--12, 2021.

\bibitem[Yang et~al.(2021)Yang, Xing, Xia, Chen, Ye, and Li]{yang2021don}
Yang, B., Xing, Z., Xia, X., Chen, C., Ye, D., and Li, S.
\newblock Don’t do that! hunting down visual design smells in complex uis
  against design guidelines.
\newblock In \emph{2021 IEEE/ACM 43rd International Conference on Software
  Engineering (ICSE)}, pp.\  761--772. IEEE, 2021.

\end{thebibliography}
